\documentclass[]{article}
\newcommand{\bb}{\begin{eqnarray}}
\newcommand{\ee}{\end{eqnarray}}
\begin{document}
%\title{Factors of two in Hawking temperature?}
\title{Regularizing tunnelling calculations of Hawking temperature}
\author{
Bhramar Chatterjee %\thanks{bhramar.chatterjee@saha.ac.in}~
and P. Mitra\\%\thanks{parthasarathi.mitra@saha.ac.in}\\
Saha Institute of Nuclear Physics\\
Block AF, Bidhannagar\\
Calcutta 700 064}
\date{}
%\date{1107.1336}
\maketitle
\begin{abstract}
Attempts to understand
Hawking radiation as tunnelling across a black hole horizon
require the consideration of singular integrals. Although
Schwarzschild coordinates lead to the standard Hawking temperature,
isotropic radial coordinates may appear to produce an incorrect value.
It is demonstrated here how the proper regularization of singular integrals
leads to the standard temperature
for the isotropic radial coordinates as well as for other smooth
transformations of the radial variable,
which of course describe the same black hole.
\end{abstract}
%\pacs{04.70.Dy}
\newpage%\bigskip

A classical black hole possesses a horizon beyond which nothing
can escape. But there is a relation involving the area
of the horizon and the mass of the black hole 
bearing a close similarity  with
the laws of thermodynamics, allowing the definition of an entropy
and a temperature \cite{Bek}. 
This analogy was understood to be of quantum origin 
after the theoretical prediction of radiation from black holes \cite{Hawk}. 
For a Schwarzschild black hole, the radiation, which is thermal, has a
temperature
\bb
T_H={\hbar\over 8\pi M},
\ee
where $M$ is the mass of the black hole.

Attempts to understand the emission of particles across the horizon as a 
quantum mechanical tunnelling process \cite{ruffini,PW,zerbini}
were initially successful, but soon yielded mixed results.  For instance,
it was pointed out \cite{anti} that this approach seems to produce a
temperature that is {\it double} the standard value $T_H$
if standard Schwarzschild coordinates are used. 
This was then explained \cite{PM} as being due to a neglect of
boundary conditions. The standard metric can represent
either a black hole or a white hole. It is necessary to distinguish between
the two by selecting boundary conditions. If this is done, there is no
problem with the value of the temperature. 

Of course, the correct result
can be obtained more convincingly by using coordinates like the
Painlev\'{e} coordinates which are nonsingular across the horizon, 
and can distinguish between black holes and white holes \cite{CGM}, but the
Schwarzschild coordinates are more familiar, so it is always useful to
understand what happens in these coordinates.

However, the use of another set of
singular coordinates, namely the isotropic coordinates, has been fraught with
problems. While these coordinates, along with
Schwarzschild coordinates, were once
argued to lead to {\it double} the accepted Hawking temperature \cite{anti},
the correct value has been reproduced in some studies,
for example \cite{jing}, but some others, for example \cite{vanzo},
find {\it half} the accepted value of the Hawking temperature.
The discrepancy arises mainly from
the choice of contours for some singular integrals involved.
We shall study the coordinates more generally 
by considering smooth transformations of the radial coordinate of the 
Schwarzschild black hole. The singular integrals can be consistently
evaluated by considering appropriately {\it regularized} functions of the
radial coordinates.

A particle, taken massless for simplicity,
is described in the Schwarzschild background 
\bb
ds^2=-(1-\frac{2M}{r})dt^2+(1-\frac{2M}{r})^{-1}dr^2+r^2d\Omega^2
\ee
by the Klein-Gordon equation
\bb
\hbar^2(-g)^{-1/2}\partial_\mu(g^{\mu\nu}(-g)^{1/2}\partial_\nu\phi)=0.
\ee
One sets
\bb
\phi=\exp(-{i\over\hbar}S)
\ee
and obtains to leading order in $\hbar$ the equation
\bb
g^{\mu\nu}\partial_\mu S\partial_\nu S=0.
\ee
Separation of variables suggests
\bb
S=Et+C+S_0(r),
\ee
where $E$ is the energy, $C$ is a constant and  angular coordinates are
neglected. The equation for $S_0$ becomes
\bb
-{E^2\over 1-\frac{2M}{r}} + (1-\frac{2M}{r})S_0'(r)^2=0
\ee
in the Schwarzschild metric. The solution of this equation is
\bb
S_0(r)=\pm E\int^r{dr\over  1-\frac{2M}{r}}.
\ee
The two signs correspond to the
fact that there can be incoming/outgoing solutions. The
singularity at the horizon $r=2M$ is usually regularized
by changing $r-2M$ to  $r-2M-i\epsilon$. In the limit,
\bb
\frac{1}{r-2M-i0}=P\frac{1}{r-2M}+i\pi\delta(r-2M),
\ee
where $P()$ stands for the principal value. This produces an
imaginary part of $S_0$ which survives in $|\phi|$:
\bb
S_0(r)=\pm 2ME[i\pi+ {\rm real~part}].
\ee
%For the outgoing solution (the negative sign),
%the imaginary part would na\"{i}vely seem to
%yield a decay factor $\exp (-\frac{2\pi ME}{\hbar})$ in the amplitude
%and hence a factor $\exp (-\frac{4\pi ME}{\hbar})$ in the probability,
%signalling a temperature \cite{anti}
%\bb
%{\hbar\over 4\pi M}=2T_H,
%\ee
%twice as big as the standard Hawking temperature.
%
One must take into account the incoming solution as well as the outgoing one:
\bb
S_{in/out}=Et+C\pm E[2M\cdot i\pi+ {\rm real~part}].
\ee
One approach here is to determine $C$ so as to cancel the 
imaginary part of $S_{in}$ to ensure that the classical incoming probability
is unity, as is appropriate for a black hole \cite{PM}. Thus,
\bb
C=-2i\pi ME,\quad
S_{out}=Et-E[4M\cdot i\pi+ {\rm real~part}],
\ee
implying a decay factor $\exp (-\frac{4\pi ME}{\hbar})$ in the amplitude,
and a factor $\exp (-\frac{8\pi ME}{\hbar})$ in the probability,
in agreement with the standard value of the Hawking temperature.
%There are other approaches too, which lead to this value from the above $S_0$.

These calculations are in the familiar Schwarzschild coordinates.
One may instead use isotropic coordinates,
\bb
ds^2=-{(1-\frac{M}{2\rho})^2\over (1+\frac{M}{2\rho})^2}dt^2+
(1+\frac{M}{2\rho})^4(d\rho^2+\rho^2d\Omega^2).
\ee
This form of the metric is obtained by rewriting
\bb
r=\rho+M+{M^2\over 4\rho}=2M+{(\rho-{M\over2})^2\over\rho}.
\label{rho}\ee
There is no change in $t$ or the angular coordinates,
but $r$ is replaced by the new variable $\rho$.
Clearly, the horizon is at $\rho=M/2$. In general, one has to solve a
quadratic to find $\rho$ for a particular $r$, and there are two solutions,
which become complex for $0<r<2M$.

In this case, the radial part of the solution is
\bb
S_0=\pm E\int^\rho d\rho {(1+\frac{M}{2\rho})^3\over 1-\frac{M}{2\rho}},
\ee
which can be written near the horizon $\rho\approx M/2$ as
\bb
S_0=\pm 4ME\int^\rho {d\rho \over \rho-\frac{M}{2}}.
\ee
The appearance of $4ME$, instead of the $2ME$ which arose
in Schwarzschild coordinates, may suggest that 
the contribution to the imaginary part of $S$ would be double the contribution
coming in Schwarzschild coordinates, leading
to a temperature $\frac{1}{16\pi M}$, half of the standard value \cite{vanzo}.

However, one must realize that the transformation to isotropic coordinates 
does not change the
black hole and hence cannot alter the temperature, which depends only on the
mass $M$ of the black hole. There must be
some way of modifying the above calculation, as we explain now.

Note that the relation (\ref{rho}) between $r$ and $\rho$ implies that
\bb
\frac{dr}{r-2M}=2\frac{d\rho}{\rho-M/2}-\frac{d\rho}{\rho}.
\ee
The last term is regular at the horizon, and cannot provide
any imaginary piece, so that any imaginary
contribution of $\frac{dr}{r-2M}$ must be
twice as much as that of $\frac{d\rho}{\rho-M/2}$.
Both cannot yield the value $i\pi$ on integration if consistently regularized.

As is widely known (cf. \cite{anti,jing}), some radial
variables behave differently from the standard $r$.
Inside the horizon, $\rho$ becomes
complex, as indicated above, because of the square root relationship
in the transformation equation (\ref{rho}).
As a result, the path across the horizon involves a change of $\pi/2$ instead
of $\pi$ in the phase of the complex variable $\rho-M/2$. This produces
a factor $i\pi/2$. However, 
as changes of contour have been criticized \cite{vanzo}, 
we shall work out other approaches. 

Note first that for imaginary $x$, there is
no imaginary contribution from $\frac{dx}{x-i\epsilon}$, while for real $x$ in
\bb
\frac{1}{x-i\epsilon}
= \frac{x}{x^2+\epsilon^2} + \frac{i\epsilon}{x^2+\epsilon^2},
\ee
the first term on the right is regular and real in the limit $\epsilon\to0$
and the second term imaginary. On integrating the second term 
from negative values to positive ones one would obtain 
$i\tan^{-1}(x/\epsilon)|_{-\infty}^\infty=i\pi$ in the limit, 
but now the integration is from zero to positive values and one obtains  
\bb
\int_0^\infty dx \frac{i\epsilon}{x^2+\epsilon^2}=
i\tan^{-1}(x/\epsilon)|_0^\infty=i\pi/2.
\ee 
%Another way of seeing this is to consider the generalized function
%\bb
%\frac{1}{(x-i0)^\frac12}
%= (x-i0)^\frac12[P\frac{1}{x}+i\pi\delta(x)]
%= (x-i0)^\frac12 P\frac{1}{x}+i(\pi/2)\delta(\sqrt{x}).
%\label{1/2}\ee
The reduced factor $i\pi/2$, together with $4ME$, yield $2iME\pi$ exactly 
as before, and the temperature becomes $\frac{1}{8\pi M}$ again.

We shall now consider the regularization of the singular integral
not just for isotropic coordinates, but for a 
more general transformation from $r$ to $R$,
\bb
r=r(R).
\ee
%which has to be a smooth function. 
The radial part is now written as $S_0(R)$. 
It satisfies the equation
\bb
-{E^2\over 1-\frac{2M}{r(R)}} + {(1-\frac{2M}{r(R)})\over r'(R)^2}S_0'(R)^2=0
\ee
This yields
\bb
S_0(R)=\pm E\int^R {r'(R)dR\over 1-\frac{2M}{r(R)}},
\ee
which can be written as 
\bb
S_0(R)=\pm E\int^{r(R)} {dr(R)\over 1-\frac{2M}{r(R)}}=S_0(r(R)),
\label{inv}\ee
showing the formal invariance of $S_0$ and hence of the temperature derived 
from it.

It is instructive to express the right hand side in terms of $R$.
Let $R_0$ be the value of $R$ at the horizon:
\bb
2M=r(R_0).
\ee
Continuity requires that $r\to 2M$ as $R\to R_0$.
A large class of transformations which satisfy such a condition have
\bb
r-2M\approx C(R-R_0)^\alpha
\label{alpha}\ee
near the horizon, where $C,\alpha$ are non-vanishing constants. 
If $r(R)$ is a smooth function near the horizon and
the $n^{\rm th}$ derivative of $r(R)$ is the lowest with a 
non-vanishing value at $R_0$, 
$\alpha=n$ and $C$ is proportional to that derivative.
%If $R(r)$ is a smooth function near the horizon and
%the $n^{\rm th}$ derivative of $R(r)$ is the lowest with a 
%non-vanishing value at $2M$, 
%$\alpha=1/n$ and $1/C^{1/\alpha}$ is proportional to that derivative.
For simple transformations, $\alpha$ may be unity, but for isotropic
coordinates, $\alpha=2$. It is seen that near the horizon
\bb
r'(R)\approx\alpha C(R-R_0)^{\alpha-1}.
\ee
Combining these, we obtain, near the horizon,
\bb
S_0(R)\approx \pm 2ME \alpha\int^R {dR\over R-R_0}.
\ee
The extra factor $\alpha$ is the cause of the confusion, suggesting that the
temperature changes when an $R$ coordinate with $\alpha\neq 1$
is used. But recalling the alternative form (\ref{inv}), one also sees that
near the horizon
\bb
2ME\int^{r(R)} {dr\over r-2M}\approx  2ME \alpha\int^R {dR\over R-R_0},
\ee
indicating that the factor $\alpha$ has to get absorbed in the $R$ integral,
which therefore has to possess a factor $1/\alpha$.
This was explicitly shown above for $\alpha=2$.
More generally, one sees from (\ref{alpha}) that
\bb
\frac{dr}{r-2M}\approx\alpha\frac{dR}{R-R_0},
\ee
so that the problem factor $\alpha$ formally gets absorbed by the integral.
One has to regularize this by introducing $i\epsilon$ once again.
If $r(R)$ is a smooth function, so that $\alpha=n$, one may write, for $x=r-2M$,
\bb
\int dx^\frac{1}{n}\frac{1}{x^\frac{1}{n}}
\stackrel{def}{=}\int dx^\frac{1}{n} x^\frac{n-1}{n}\frac{1}{x-i0}
&=& \int dx^\frac{1}{n} x^\frac{n-1}{n}[P\frac{1}{x}+i\pi\delta(x)]\nonumber\\
&=&(\frac{1}{n})\int dx[P\frac{1}{x}+i\pi\delta(x)].
\ee
%If $R(r)$ is smooth, $\alpha=1/n$ and (\ref{1/2}) generalizes to
%\bb
%\frac{1}{(x-i0)^n}
%= P \frac{1}{x^n} +i\pi (-1)^{n-1}\delta^{(n-1)}(x)/(n-1)!
%= P \frac{1}{x^n} +i\pi n\delta(x^n),
%\ee
%where $\delta^{(n-1)}$ stands for the $(n-1)^{\rm th}$ derivative of $\delta$.
Thus the factor $1/\alpha=1/n$ required for invariance
under the coordinate transformation is explicitly achieved with a suitable
definition of the singular integral.
It may be noted that the generalized function $\frac{1}{(x-i0)^{1/n}}$,
although well defined, does not contain a delta function for $n>1$ and
is not appropriate.

In short, there is no problem with the calculation of the Hawking temperature
of a black hole through the use of redefined radial coordinates
like isotropic coordinates.
If one describes a black hole by transforming $r$ smoothly,
the proper regularization of the singular integral confirms the
invariance of $S_0$ and hence the Hawking temperature.

\bigskip

We thank Amit Ghosh for discussions.

\end{document}